\documentclass[twocolumn,showpacs,preprintnumbers,amsmath,amssymb,aps]{revtex4}
\input epsf
\usepackage{graphicx}

\begin{document}

\title{
Heterogeneities and Topological Defects in Two-Dimensional Pinned Liquids} 

\author{J.-X. Lin$^{1,2}$, C. Reichhardt$^{1}$, 
Z.~Nussinov$^{1,3}$, Leonid P. Pryadko$^{2}$, and 
C.J. Olson Reichhardt$^{1}$}
\affiliation{ 
{$^1$}Center for Nonlinear Studies and Theoretical 
Division, 
Los Alamos National Laboratory, Los Alamos, New Mexico 87545\\
{$^2$}Department of Physics, 
University of California, Riverside, California 92521\\
{$^3$}Department of Physics, Washington University, St. Louis, Missouri
63130} 

\date{\today}

\begin{abstract}
We simulate a model of repulsively interacting colloids on 
a commensurate two-dimensional triangular pinning substrate 
where the amount of heterogeneous motion that appears at melting can 
be controlled systematically by turning off a 
fraction of the 
pinning sites.  We correlate the amount of heterogeneous motion 
with the average topological defect number, time dependent 
defect fluctuations, colloid diffusion, and the form of the van Hove 
correlation function.
When the pinning sites are all off or all on, the melting occurs
in a single step. When a fraction of the sites are turned off, the
melting becomes considerably broadened and signatures of a two-step
melting process appear. 
The noise power associated with fluctuations in the number of
topological defects reaches a maximum when half of the pinning
sites are removed, and the noise spectrum has a pronounced
$1/f^{\alpha}$ structure in the heterogeneous regime.
We find that regions of high mobility are associated 
with regions of high dislocation densities. 
\end{abstract}

\maketitle

\vskip2pc

\section{Introduction} 

Glassy and liquid assemblies of particles in two and three 
dimensions have been shown to exhibit dynamical heterogeneities, 
where the motion of the particles is not uniform but
occurs in correlated strings in certain regions, while
other regions are less mobile \cite{Sillescu,Ediger,Richert}. 
Numerous numerical simulations have 
found evidence of dynamical 
heterogeneities near the glass transition \cite{Donati}. 
Direct observations of correlated regions of motion have been obtained
in imaging experiments on three-dimensional (3D) 
colloidal assemblies as the system
approaches a glassy phase \cite{Blaaderen,Weeks},
and nonuniform motion has been found in polymer melts
\cite{Russell,Deschenes,Reinsberg}. 
Heterogeneous motion has also been observed directly for systems that
do not form a glassy phase but that can have a dense liquid region near 
crystallization, including   
2D colloids \cite{Marcus}  and 
dusty plasmas \cite{Lin,Goree}. 
    
In recent work on 2D systems of repulsive colloids or vortices 
which form a triangular lattice crystalline phase,
it was shown that at temperatures just 
above the melting transition, topological
defects in the form of dislocations undergo correlated annihilation and
creation, giving rise to a $1/f^{\alpha}$ noise signal in the time 
dependent dislocation density \cite{Olson}. The $1/f$ noise also 
coincides with the appearance of dynamical heterogeneities. At higher
temperatures, the dynamical heterogeneities disappear and the noise spectrum 
of the fluctuating topological defect density becomes white, 
indicating the loss of correlations.
The same system has been studied in the case where
the particles are quenched from a
high temperature liquid phase
where there is a high density 
of dislocations to a low temperature phase where the ground state is a 
triangular
lattice.  In the low temperature regime  
the dislocations created in the quench annihilate over time. The particle
motion in this annihilation process occurs
in the form of string like jumps \cite{Olson}.
This result suggests that the dynamical heterogeneities 
are directly correlated with
the motion, creation, and annihilation of topological defects.  
Further evidence that in 2D the topological defects 
are associated with dynamical heterogeneities has also been reported in the 
recent experiments of Dullens and Kegel on 2D colloidal suspensions,
where the non-sixfold-coordinated colloids were
more mobile than sixfold-coordinated colloids \cite{Kegel}.    

In a related class of systems, glassiness does not arise solely
from the particle interactions but instead occurs due to coupling
with an underlying quenched substrate.  It has already been shown in
2D systems of classical electrons in the presence of quenched
disorder that the particle motion occurs in string-like dynamical
structures where a chain of particles moves past other particles that
are pinned \cite{Wigner}.  These motions are very similar to the
string-like dynamical heterogeneities observed in systems
without quenched disorder.  
It is important to note, however, that in Ref.~\cite{Wigner}, where
the disorder was simulated as a collection of randomly located pins,
the quenched disorder had a tendency to create topological defects even at
very low temperatures \cite{Fertig}.  Therefore, the connection
between the dynamical heterogeneities and the
topological defects was not immediately apparent, and
it would be desirable to identify a system in which the amount
of heterogeneous motion could be controlled systematically.

In this work,
we propose a model of repulsively interacting particles on a substrate
in which
the disorder potential is perfectly
commensurate with the triangular crystal and therefore does not favor
the creation of topological defects.  Specifically,
we study colloidal particles interacting
via a screened Coulomb repulsion in the presence of a triangular pinning
substrate where the number of colloids is commensurate with the 
number of pinning sites.  We introduce disorder by 
shutting off a specified fraction of randomly selected pinning sites.
The crystal phase is stabilized in regions of the system that contain
a locally large density of active pins.
As the temperature is increased, the melting occurs first in regions
with higher densities of nonactive pinning sites.

The system we consider can be realized experimentally for colloids interacting
with periodic arrays of optical traps 
\cite{Grier,Bechinger,Mangold,Reichhardt,Korda,Korrda,Dholakia}. 
The melting of charged colloids interacting with triangular and square pinning 
arrays has already been studied experimentally 
\cite{Bechinger,Mangold} and 
in numerical simulations \cite{Reichhardt}.   
Experimental evidence for a coexistence of
a liquid and a solid has been obtained in a system where
colloids located at pinning sites
remain immobile while colloids in the unpinned interstitial regions
are mobile \cite{Mangold}. 
Related systems 
that can be modeled as repulsive 
particles interacting with a periodic substrate
include vortices in superconductors with artificial pinning
sites \cite{Harada,Scalettar} 
and vortices in Bose-Einstein condensates 
interacting with optical traps 
\cite{Bigelow}. 

\begin{figure}
\includegraphics[width=3.5in]{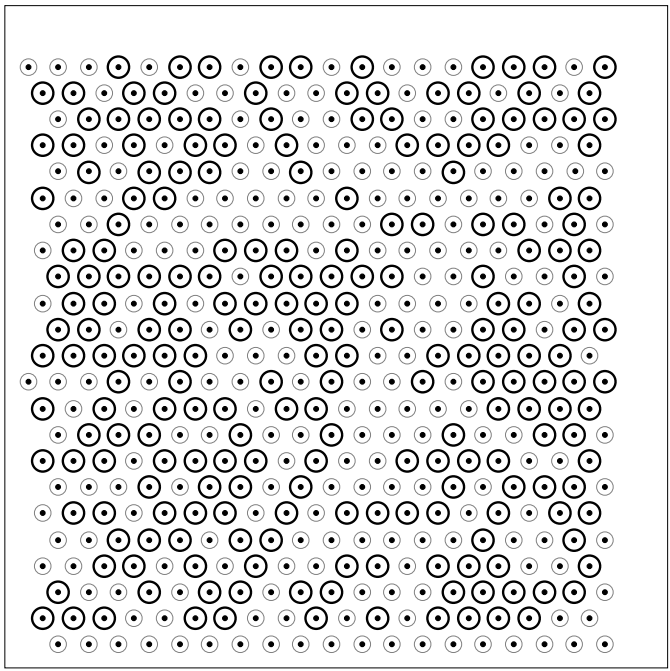}
\caption{
Location of the pinning sites (circles) and colloids (black dots). 
Large dark circles
indicate pinning sites with a finite $f_{p}$ while small gray 
circles indicate pins that have been shut off by setting
$f_{p} = 0.0$. In this image, the fraction of sites with finite $f_{p}$ is 
$n_p = 0.5$. 
}
\end{figure}

\begin{figure}
  \includegraphics[width=3.5in]{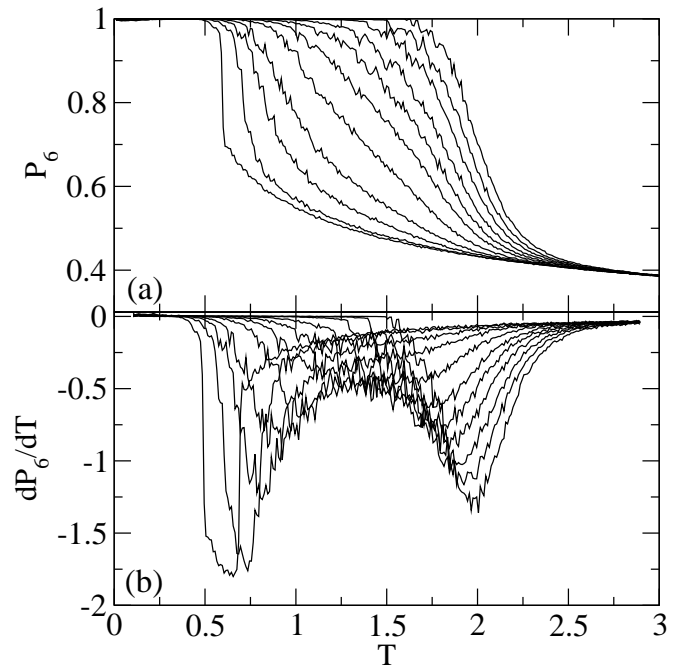}
\caption{
(a) The fraction of colloids with six-fold coordination number $P_{6}$ vs 
$T$ for various $n_{p}$. From right to left, $n_p = 1.0$, 0.9, 0.8, 0.7,
0.6, 0.5, 0.4, 0.3, 0.2, 0.1, and 0.
(b) The corresponding $dP_{6}/dT$ curves.      
}
\end{figure}

\begin{figure}
\includegraphics[width=3.5in]{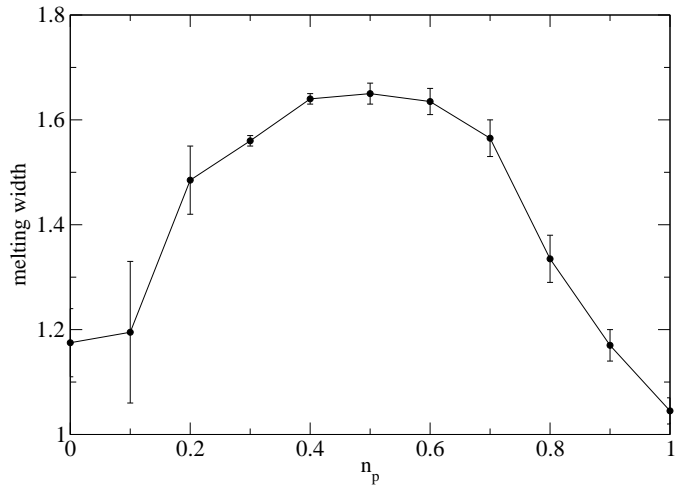}
\caption{
The width of the melting curve 
versus pinning fraction $n_p$ obtained for the system in Fig.~2. 
}
\end{figure}

\begin{figure*}
\includegraphics[width=5.0in]{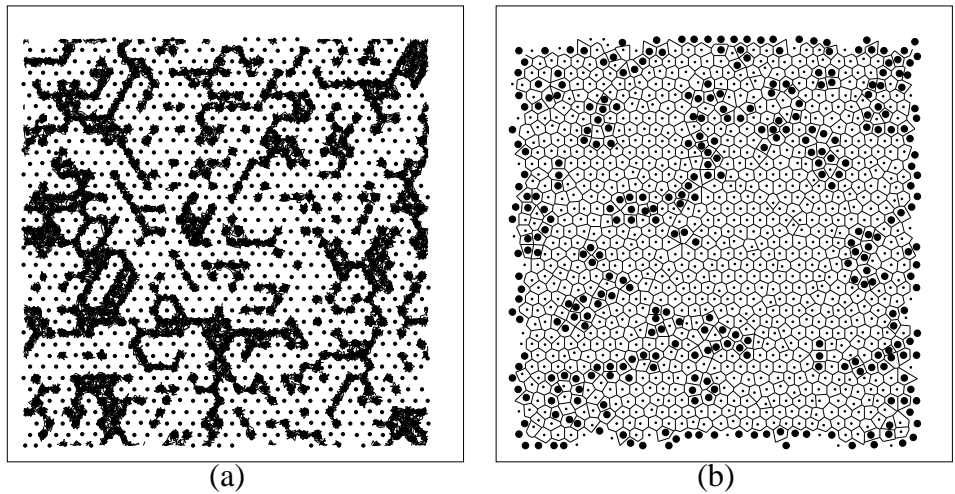}
\caption{
(a) Colloidal trajectories (black lines) and colloidal positions
(black dots) over a fixed period of time in a system with $n_{p} = 0.5$ at
$T = 1.5$. (b) Voronoi plot of the system in (a) for 
a single frame. Small black dots indicate sixfold-coordinated particles
while large black dots indicate particles
with coordination numbers other than six.     
}
\end{figure*}

\section{Numerical Simulation Method and Parameters}

We model a 2D system of a monodisperse
assembly of $N$ colloids using a Brownian dynamics simulation   
with periodic boundary conditions 
in the $x$ and $y$ directions. 
The equations of motion for the colloidal particles are overdamped and 
we neglect hydrodynamic interactions, which is a reasonable assumption
for charged particles in the low volume fraction limit.
A single colloid $i$ 
obeys the overdamped equation of motion 
\begin{equation}
\eta\frac{ d{\bf R}_{i}}{dt} = {\bf F}_{i}^{cc} + {\bf F}^{T}_{i} + 
{\bf F}^{s}_{i}  
\end{equation}
Here $\eta$ is the damping constant which is set to
unity. The colloid-colloid interaction force is 
${\bf F}_{i}^{cc} = -q_{i}\sum^{N}_{i\neq j}\nabla_i V(r_{ij})$,
where the colloid-colloid interaction potential is a screened Coulomb 
interaction of the form
\begin{equation} 
V(r_{ij}) = (q_{j}/|{\bf r}_{i} - {\bf r}_{j}|)\exp(-\kappa|{\bf r}_{i} 
- {\bf r}_{j}|). 
\end{equation}  
Here $q_{j(i)}$ is the  charge on particle $j$($i$), $\kappa$ is
the inverse screening length which is set to $3/a$,
and ${\bf r}_{i(j)}$ is the position of particle
$i$($j$).
Throughout this study 
the density of colloids is kept fixed at $n_{c} = 1.0$ which
gives a colloid lattice constant of $a = 1.0$.  
We also fix the system size to $L=24$.
Because the colloid-colloid interaction is screened, at long distances the
force between two colloids is negligible; thus, we place a cutoff 
on the interaction at $5a$. For larger cutoffs
we find no change in the results.  
The thermal force ${\bf F}^{T}$ is modeled as
random Langevin kicks 
with the properties $\langle{\bf F}^{T}_{i}\rangle = 0$ and
$\langle{\bf F}^{T}(t){\bf F}^{T}(t^{\prime})\rangle 
= 2\pi\eta k_{B}T\delta(t - t^{\prime})$. 
The pinning comes from the substrate force
${\bf F}^{s}_{i}$. The pinning sites are modeled as parabolic traps of radius 
$r_{p}$ and
maximum strength $f_{p}=2.0$ 
which are placed in a triangular array with
lattice constant $a$, commensurate with the colloidal lattice. 
As an initial condition, each colloid is placed inside a pinning site.
We turn off some of the pins by setting $f_p=0$ at some sites, keeping
$f_p$ finite at a fraction $n_p$ of randomly chosen ``active'' pinning sites.
We gradually increase the temperature up to 3.0 in increments
of $0.01$. The melting transition is identified by examining the
density of topological defects and the diffusion. 
A clean system with all the pinning turned off 
melts at $T = 0.6$.  

\section{Topological Defects and Noise}
\subsection{Defect Density and Melting}

In Fig.~1 we illustrate a system with $n_p=0.5$ where 
we have turned off half of the pinning sites 
by setting $f_p=0$ at these sites.
At $T = 0$ the colloids form a triangular lattice that is commensurate with 
the substrate. 
To determine the melting 
behavior as a function of temperature we measure 
the density of sixfold coordinated particles $P_{6}$
using
a Voronoi or Wigner-Seitz construction. 
For a perfectly triangular lattice, $P_{6} = 1.0$.
Topological
defects such as dislocations produce $5-$ and $7-$fold coordinated particles. 
In Fig.~2(a) we plot $P_{6}$ vs $T$ for systems with
varied pinning fractions of $n_{p} = 0$ (no pins active) to $1.0$ 
(all pins active).
When thermally induced defects begin to appear, $P_{6}$ drops.  
As $n_{p}$ increases, the drop in $P_{6}$ shifts to higher temperatures.
The sharpest drop in $P_{6}$ occurs at $n_{p} = 0$ where none of the
pins are active.  The drop becomes steeper again as $n_{p}$  
approaches 1 where all the pins are active.
For intermediate fillings the drop in $P_{6}$ is broadened.
At high temperatures $T >  2.5$, all of the $P_{6}$ curves
come together near a value of $P_{6} = 0.4$.      

In Fig.~2(b) we plot the derivatives of the curves in Fig.~2(a). 
For intermediate filling $0.4 < n_{p} < 0.75$, there are two dips
in $dP_{6}/dT$. The first dip at lower temperatures 
is associated with the onset of 
dislocations among the colloids in the unpinned regions, while
the second dip corresponds to the onset of dislocations in the pinned regions. 
For $n_{p} < 0.4$, 
the first dip near $T = 0.65$ is the sharpest and the dip
shifts to higher temperatures as $n_{p}$ increases. The
magnitude of the first dip decreases until $n_{p} = 0.5$, 
and for $n_{p} > 0.5$ the
first dip is lost. 
At $n_{p} = 0.5$ a small shoulder starts to occur in $dP_{6}/dT$
near $T = 1.75$, indicating the onset of the second dip. 
The second dip continues
to grow in magnitude and shifts to higher $T$ with increasing
$n_p$ until at $n_{p} = 1.0$ the dip occurs at 
$T = 2.0$. For higher temperatures
$T>2.5$, all of the $dP_6/dT$ curves merge. 

\begin{figure*}
\includegraphics[width=5.0in]{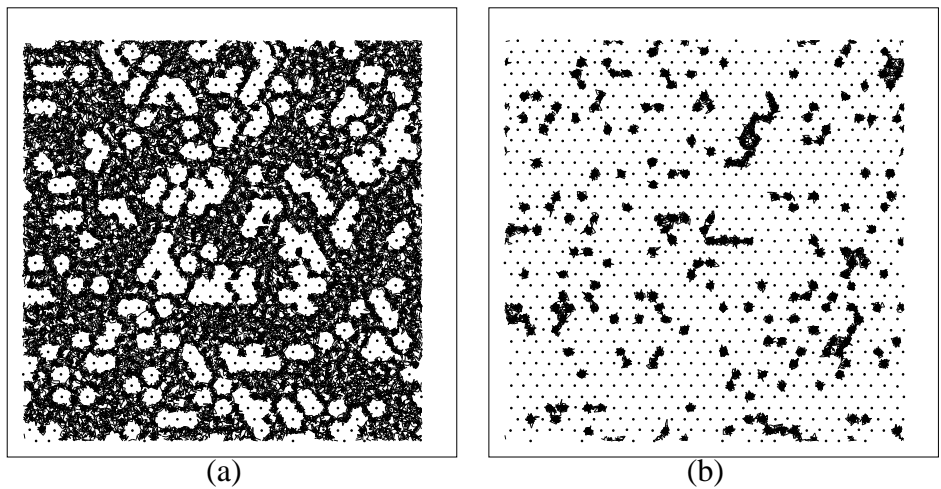}
\caption{
Colloidal trajectories (black lines) and positions (black dots) for a system 
at fixed $T = 1.5$ for (a) $n_{p} = 0.2$ and (b) $n_{p} = 0.8$. 
}
\end{figure*}

The data in Fig.~2(b) indicate that there 
are two characteristic disordering regimes.  The first coincides 
with the temperature
at which the particles located in the unpinned regions effectively melt, 
and the second corresponds to the temperature at which the particles
in the pinned regions melt. From the curves in Fig.~2 the width of each 
melting transition can be determined by measuring the distance from
the beginning of the dip in $dP_{6}/dT$ to the temperature where
$dP_6/dT$ begins to saturate. 
In Fig.~3 we plot the melting width vs $n_{p}$, showing that the width reaches
a maximum value at $n_{p} = 0.5$.  
The width is smaller for $n_{p} = 1.0$ than for the unpinned
case of $n_{p} = 0$.     
     
The system
at $n_{p} = 0.5$ and $T = 1.5$ can be 
regarded as a mixture of a solid pinned
phase and a liquid phase, and thus the motion and diffusion of 
particles is highly heterogeneous. At 
$n_{p} = 0$ and $T=1.5$ in Fig.~2(a),
$P_{6}=0.5$, indicating that at this temperature the system is in 
a strongly disordered liquid state. 
For the same $T=1.5$ at $n_p = 1.0$, $P_{6} = 1.0$, 
indicating that the system is
a completely triangular solid. In contrast, for $n_{p} = 0.4$ at $T = 1.5$,
$P_{6} = 0.67$.  In this case, it would be
expected that the colloids located at pinning sites that have been turned
off should have a liquid like behavior, while
the colloids at the active pinning sites should behave like
a solid. The dislocations
and fluctuations in the dislocation density should then 
be associated with the liquidlike unpinned regions. 

In Fig.~4(a) we plot the colloid trajectories (black lines) and 
colloidal positions (black dots) for a fixed period of time for 
a system with $n_{p} = 0.5$ and  $T = 1.5$, showing
a highly heterogeneous motion of particles in correlated groups where
the pinning is deactivated.
In Fig.~4(b) we illustrate the corresponding
Voronoi construction for a single frame indicating the locations of the
topological defects as nonsixfold coordinated particles. 
In general, the dislocations are
located in the {\it same} regions where the correlated particle motions 
are occurring. 

The amount of heterogeneous motion that appears depends on both the 
filling fraction $n_p$ and the
temperature. In general, for high temperatures at 
all fillings, the motion is homogeneous and
the defect density is high. 
In Fig.~5(a) we show the trajectories for the case of 
$n_{p} = 0.2$ at $T=1.5$ where the system is mostly in the liquid state 
with a small number of
pinned colloids. In Fig.~5(b) we plot the trajectories 
for the same temperature at 
$n_{p} = 0.8$, where most of the system is pinned and 
a small number of colloids show
extra motion at their sites but do not change neighbors.  

\subsection{Fluctuations in the Defect Density and Noise} 

We next consider how the filling fraction affects the time
dependent fluctuations of topological defect density. 
In Fig.~6 we plot the time series of the density of
six-fold coordinated particles, $P_{6}(t)$, in a system with $n_p=0.5$
for $T = 1.5$ (upper curve) and $T = 4.0$ (lower curve).
For $T = 4.0$, the system is strongly disordered with 
$\langle P_{6}\rangle = 0.36$, 
while at  $T = 1.5$ the system is in the heterogeneous phase and
$P_6$ varies from 0.86 to 0.68. 
The fluctuations in the defect density are much more rapid for $T = 4.0$
than at $T = 1.5$.
For much longer time
series at $T=1.5$, there are additional long time 
fluctuations with $P_{6}$ rising as high as $0.9$ on
occasion. 
In order to quantify the fluctuations in the defect density, we compute 
the power
spectrum of the time series, 
\begin{equation}
S(f) = \left |\int P_6(t)e^{-2\pi if t}dt\right|^2 .
\end{equation} 
The noise power $S_{0}$ is the value of 
$S(f)$ 
averaged over a specific 
frequency octave.   
Fig.~7(a) shows 
$S(f)$
for $T = 1.5$ where a clear $1/f^{\alpha}$ signal 
appears with $\alpha = 1.45$. In Fig.~7(b) we plot
$S(f)$ 
for the same system at $T = 4.0$ where the noise spectrum is 
closer to white  
with $\alpha = 0.1$. 
For intermediate temperatures $\alpha$ gradually changes from $1.45$ to $0.1$
or a white spectrum, indicating that there is little or no correlation in the
defect fluctuations at the higher temperatures.
We note that for the case where there is no pinning, previous studies found
dynamical heterogeneities occurring just above
the melting transition \cite{Olson}. 
In these studies, similar $1/f^{\alpha}$ 
dislocation density noise appeared in this
regime; however, the value of 
$\alpha$ had a maximum of $1.0$. 
The larger value of $\alpha$ that we observe here 
implies that for the pinned system, there are stronger
correlations in the annihilation and creation of the 
dislocations compared to the unpinned system. 
In Fig.~8 we plot the noise power $S_{0}$ vs $T$ for systems with
$n_{p} = 0$ (squares) and $n_{p} = 0.5$ (circles).  
The clean system shows a maximum noise power at
$T = 0.75$ and then a slow drop in noise power for higher temperatures 
as observed in previous simulations \cite{Olson}.
For the case of $n_{p}=0.5$, the noise power begins to increase
near $T = 1.0$ and reaches a much higher maximum near
$T = 1.8$ before decreasing at higher temperatures. 
The noise power for the different 
fillings becomes equal near $T = 2.5$, which is also the temperature
at which the $dP_{6}/dT$ curves merge in Fig.~2(b). 
For higher filling fractions, the peak
in the noise power shifts to higher $T$ and there is a 
slight decrease in the maximum value of 
the noise power. Although the width of the noise power 
peak is broadened for the 
$n_{p} = 0.5$ case compared to the $n_p=0$ case,
we do not observe two peaks, which would be indicative 
of two melting transitions.
  
\begin{figure}
\includegraphics[width=3.5in]{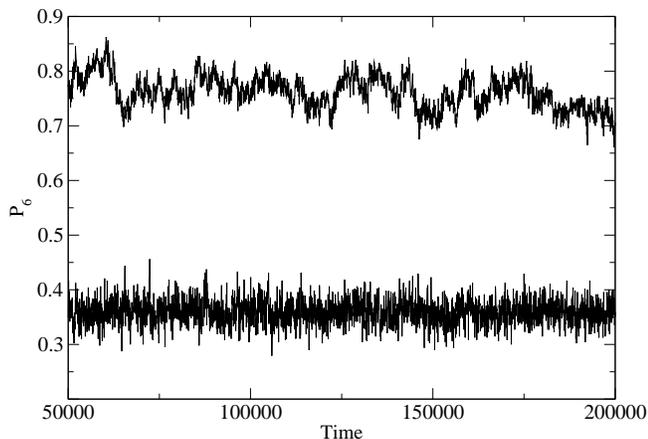}
\caption{Time series of the fraction of colloids with six-fold coordination,
$P_6$,
for a system with $n_{p} = 0.5$.
Top curve, $T = 1.5$; lower curve, $T = 4.0$.
}
\end{figure}

\begin{figure}
  \includegraphics[width=3.5in]{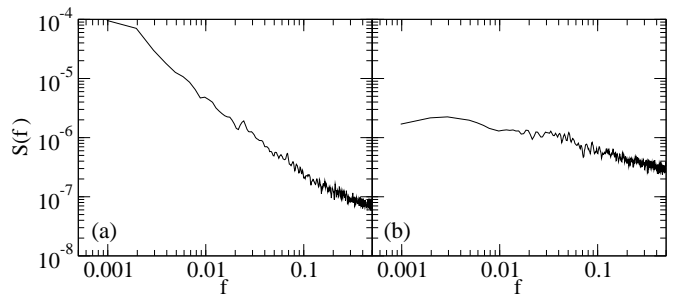}
\caption{Power spectrum 
$S(f)$ 
for the same system in Fig.~6 at  
(a) $T = 1.5$ and (b) $T = 4.0$. 
}
\end{figure}

\begin{figure}
\includegraphics[width=3.5in]{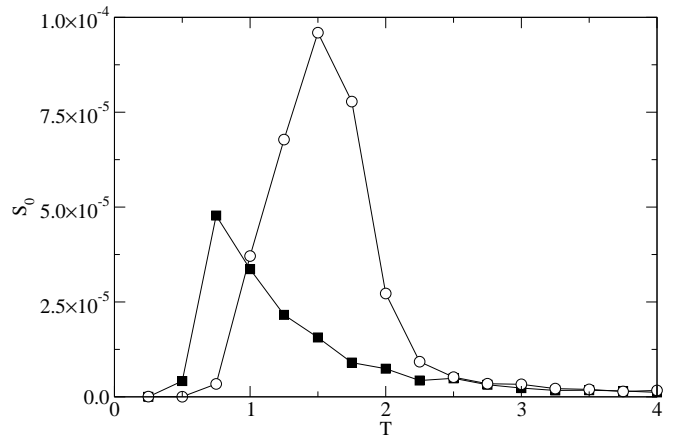}
\caption{
Noise power $S_{0}$ vs $T$ for a system with 
$n_{p} = 0.0$ (squares) and $n_{p} = 0.5$ (circles).
}
\end{figure}

In Fig.~9 we show how $\alpha$ evolves as a function of temperature for a
system with $n_{p} = 0.5$  
as obtained from the power spectrum. The peak value
of $\alpha$ occurs at a temperature of $T = 1.5$ which 
also corresponds to the peak in
the noise power $S_{0}$ in Fig.~8. As $T$ increases, there is a slow fall off 
to a white noise spectrum with $\alpha \approx 0$ at $T = 4.0$. 
This result shows that 
in the regime where the motion is highly heterogeneous, the noise spectrum
is broad, indicating
correlations in the creation and destruction of the topological defects.
At the high temperatures where the motion is uniform, the correlation in the
noise spectrum is lost.  

\begin{figure}
\includegraphics[width=3.5in]{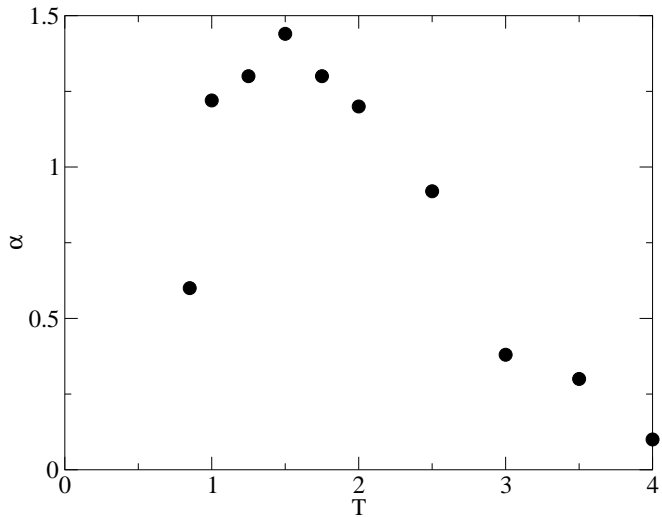}
\caption{The exponent $\alpha$, obtained from the power spectrum, vs $T$ for 
the same system in Fig.~6 with $n_{p} = 0.5$.
}
\end{figure}

\subsection{Effect of Substrate Strength}

We next consider the effects of varying the pinning strength of the substrate. 
In the previous analysis the pinning strength was fixed at 
$f_{p} = 2.0$ so that there was a clear distinction between the
melting of the pinned species and of the unpinned species. 
When $f_{p}$ is varied, 
the width of the heterogeneous regime as a function of $T$ can be 
increased or decreased. We perform
a series of simulations with fixed $n_{p} = 0.5$ and varied $f_{p}$. 
In Fig.~10 we plot 
the temperature at which 
the system enters the liquid phase (squares) vs $f_p$. 
This transition is defined as the point where $P_{6}$ 
saturates to a value near $0.4$. 
We also plot the temperature at which
the first topological defects appear (circles) vs $f_{p}$. 
For $f_{p} > 0.4$, the temperature at which defects first appear
saturates at $T\approx 1.14$, while the transition into the liquid
phase shifts to higher temperatures with increasing $f_p$.
The saturation of the dislocation onset line delineates the
crossover to the strong pinning limit, and indicates that
even though the pinning strength is increasing, the particles located in the
non-pinned regions melt at a constant temperature.
The melting of the non-pinned species occurs when the thermal motion is 
strong enough to overcome
the repulsive colloid-colloid interaction force. 
Since the colloid interaction strength 
is not changing as a function of $f_{p}$, the 
unpinned colloid melting temperature saturates.   
The colloids located at active pinning sites
can only melt when the thermal fluctuations are strong enough
to enable the particles to hop out of the pinning sites. As $f_{p}$ 
is increased, the activated hopping temperature also increases. 
For $f_{p} < 0.4$, in the weak pinning regime,
the melting does not occur in a two step fashion but instead occurs
in a single step, similar to the case of 
$f_{p} = 0.0$. At $f_{p} = 0.0$ there
is still a finite window of temperature falling between the 
onset of dislocations and the saturation of $P_{6}$. 
In this case, the motion can still be heterogeneous, 
as has been previously studied; however, the dynamical 
heterogeneities for $f_p=0$ are
not located at specified regions but are moving over time so that
all the particles take part in the motion over
long times \cite{Olson}. 
This is in contrast with the finite $f_p$ case where, due to the existence
of pinning sites, only certain particles take part in the motion.   

\begin{figure}
\includegraphics[width=3.5in]{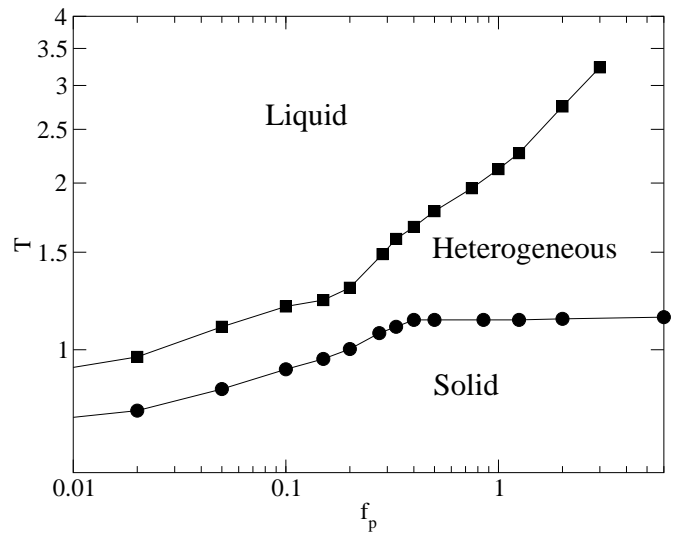}
\caption{
Diagram of the different regimes 
for $T$ vs pinning strength $f_{p}$ at fixed  
$n_{p} = 0.5$. Circles indicate the onset of the first topological
defects. Squares show the temperature at which 
$P_{6}$ saturates to a value of $0.4$. 
}
\end{figure}

An interesting effect indicated by Fig.~10 is that, up to $f_{p} = 0.4$, 
pinning can effectively {\it increase} 
the melting temperature
of the entire lattice by up to $\Delta T = 0.3$, 
while for $f_{p} > 0.4$, the two step melting scenario occurs. 
This increase in melting temperature for weak pinning occurs 
since even at $n_{p} = 0.5$, the triangular colloidal lattice
is still commensurate with the triangular substrate. 
The thermal fluctuations that the colloids experience
originate both from the applied thermal Langevin force and also 
from the disordered motion of the
surrounding colloids.  Since the colloids that are at the pinning sites 
are more constrained, they fluctuate less than unpinned colloids. 
This leads to an overall reduced fluctuating force on all the colloids
and results in an increase of the melting temperature.
Once the first dislocations appear, the pinning in this regime is not strong
enough to create the two step melting found above $f_p=0.4$. 

\section{Diffusion Measures} 
\subsection{Average Diffusion Coefficient}

We next consider diffusive measures. 
We calculate the distance traveled by the colloids
during a fixed period of time $\delta t=t_1-t_0$. 
The diffusion is given by
$D =  ({\bf r}(t_{1}) - {\bf r}(t_{0}))^2/\delta t$
for fixed $\delta t$. 
Based on the initial colloid positions, we can distinguish between
the initially pinned and initially
unpinned colloids, and we measure the diffusion of the two species separately. 
In Fig.~11(a) we plot $D$ vs $T$ for a system with $f_{p} = 2.0$ 
and varied $n_{p}$ for
the colloids that were initially placed in sites that had $f_p=0$.
In Fig.~11(b) we plot $D$ for the same system in Fig.~11(a) for
the particles that were initially placed at active pinning sites. 
For the unpinned colloids in Fig.~11(a),
for $n_{p} = 0.0$ at low temperatures up to $T = 0.6$, there is an 
initial increase of $D$ with increasing $T$. This is due to
the fact that the thermal forces cause the colloids to rattle inside 
the caging potential created
by the interactions with the other colloids; 
however, the system maintains triangular 
ordering and the particles do not hop out of their initial locations. 
There is a subsequent sharper increase in $D$ for $T > 0.6$ when the 
system melts and topological defects proliferate. Above the melting 
transition, $D$ 
continues to increase with $T$ as the particle can move a greater
distance during $\delta t$
at the higher temperatures. For higher values of  $n_{p}$
in Fig.~11(a), the initial 
values of $D$ for $ T < 0.6$ are monotonically shifted down, 
indicating that the pinned colloids are exerting 
a stronger confining force on the unpinned colloids. 
As $n_{p}$ is increased, the sharp increase in $D$ denoting the
melting of the unpinned particles is shifted to higher $T$, 
consistent with the measurements of $P_{6}$ from Fig.~2.
This jump broadens as $n_{p}$ increases. For fillings
of $n_{p} = 0.3$ to $0.7$, 
$D$ shows a plateau like feature followed by an additional smaller increase   
at a higher temperature which corresponds to the temperature
at which the particles in the pinning sites melt. 
The derivative of $dD/dT$ shows a single
peak at $n_{p} = 0.0$ and $0.9$ and two peak like features 
around $n_{p} = 0.5$. 
The plateaulike feature in $D$ at intermediate fillings occurs because
even though the unpinned particles have melted,
their motion is confined to the regions where there is no pinning. 
This limits the magnitude of the long time diffusion.  
Once the colloids at the pinning sites also melt, the unpinned
colloids are no longer confined and can diffuse freely.
At higher temperatures all of the diffusion curves come together. 
 
\begin{figure}
  \includegraphics[width=3.5in]{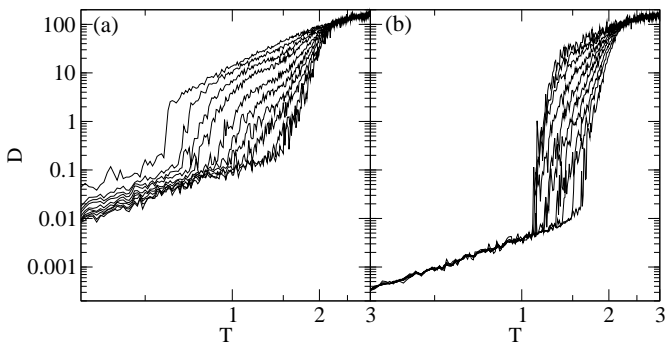}
\caption{ 
(a) Diffusion $D$ of the initially unpinned particles
vs $T$ for a system with $f_{p} = 2.0$ and
$n_{p} =$ (left) 0, 0.1, 0.2, 0.3, 0.4, 0.5, 0.6, 0.7, 0.8,
and 0.9 (right).
(b) $D$ vs $T$ for the same system in (a) for the initially
pinned colloids at
$n_{p} =$ (left) 0.1, 0.2, 0.3, 0.4, 0.5, 0.6, 0.7, 0.8, 0.9, and 1.0 (right).
}
\end{figure}

Figure 11(b) illustrates $D$ vs $T$ for the colloids that 
were initially placed in active pinning sites.
There is an initial slow increase in D   
for $T < 1.0$ which is similar to the slow increase seen
in Fig.~11(a). In this case $D$ for the pinned colloids
has a much smaller value than $D$ for the unpinned colloids.
This is due to the confining force exerted by the pinning wells
on the pinned colloids. For $n_{p} = 0.1$, the colloids at the pinning 
sites only begin to jump out of the wells 
at $T > 1.0$, in contrast to the
case in Fig.~11(a) where the jump in diffusion 
for the unpinned particles occurs at
much lower temperatures. As $n_{p}$ increases, 
the  
temperature at which the diffusion of pinned colloids begins
also increases.
This indicates that the motion of the unpinned particles 
affects the melting of the pinned particles by providing an extra
effective thermal noise.
We note that pinned colloids located near regions
of unpinned colloids do not experience the same fluctuating 
interaction force as
pinned colloids that are surrounded mostly by pinned colloids.
At high temperatures, all of the curves in Fig.~11(b) come together to the 
same value, as was the case for Fig.~11(a).   

\begin{figure}
\includegraphics[width=3.5in]{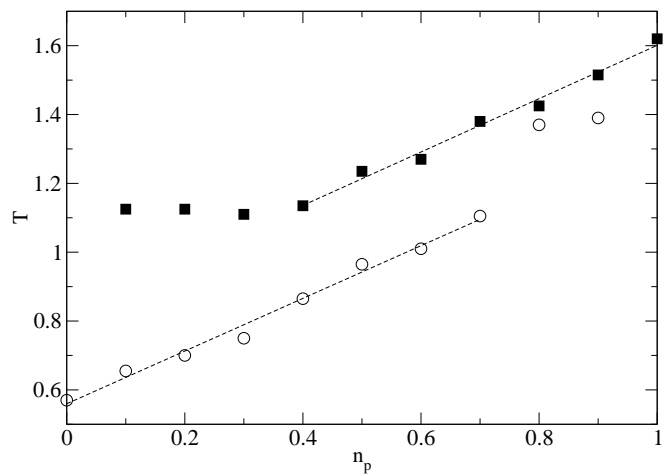}
\caption{ 
Open circles: Temperature at which the diffusion noticeably increases 
for the unpinned colloids from Fig.~11(a). 
Black squares:  Temperature at which the diffusion noticeably 
increases for the pinned
colloids from Fig.~11(b). 
Dashed lines are linear fits which both have the same slope. 
}
\end{figure}

In Fig.~12 we plot the temperature at which 
the diffusion noticeably increases for 
the pinned particles (black squares) and the unpinned particles 
(open circles) vs $n_{p}$. 
The temperature at which the unpinned colloids begin to 
diffuse significantly is lower than the temperature at 
which the pinned particles begin to diffuse.
The diffusion onset temperature for the
unpinned colloids increases linearly with 
$n_{p}$ up to $n_p=0.75$.  For $n_p>0.75$,
the diffusion onset temperature jumps to a value close to the diffusion
onset temperature for the pinned colloids.
This is due to the fact that for $n_{p} > 0.75$ most of the 
colloids at unpinned sites are completely surrounded by pinned colloids, 
and the time scale for an unpinned colloid to hop 
to another site becomes much longer.

For the pinned particles, the diffusion onset 
temperature also increases linearly with
$n_{p}$ for $n_{p} > 0.4$. 
For $n_{p} < 0.4$ the diffusion onset temperature remains fixed at a constant
value.  We note that for the lower pin fillings, the pinned colloids
are mostly surrounded by unpinned colloids. The region in which
a clearly defined two-step melting transition occurs
is $0.4 < n_{p} < 0.75$, which  
is consistent with the values of $n_{p}$ at which two peaks occur in 
$dP_{6}/dT$ and $dD/dT$.    

\subsection{van Hove correlation function} 

Another measure used to identify and 
analyze dynamical heterogeneities is the self part of the 
van Hove correlation function $G$. 
This measure gives the probability distribution that a
particle has moved a distance ${\bf r}$ during a fixed time interval $t$: 
\begin{equation}
G({\bf r},t) = N^{-1}\biggl\langle\sum^N_{i =1}\delta\biglb(
{\bf r}-{\bf r}_{i}(0) + {\bf r}_{i}(t)\bigrb)\biggr\rangle .
\end{equation} 
Systems 
with dynamical heterogeneity have non-Gaussian average
velocity distributions. 
Experimental studies have shown that in a liquid state 
without dynamical heterogeneities,
this measure produces a single Gaussian fit. In the heterogeneous phase, 
there is extra weight at larger distances and a double Gaussian fit can be 
used \cite{Kegel}.  The Gaussian distribution for the short distances
can be interpreted as corresponding to slow particles that are located in
regions with low mobility, while the second wider Gaussian distribution 
that fits the larger distances corresponds to faster particles located in
regions with higher mobility.

\begin{figure}
  \includegraphics[width=3.5in]{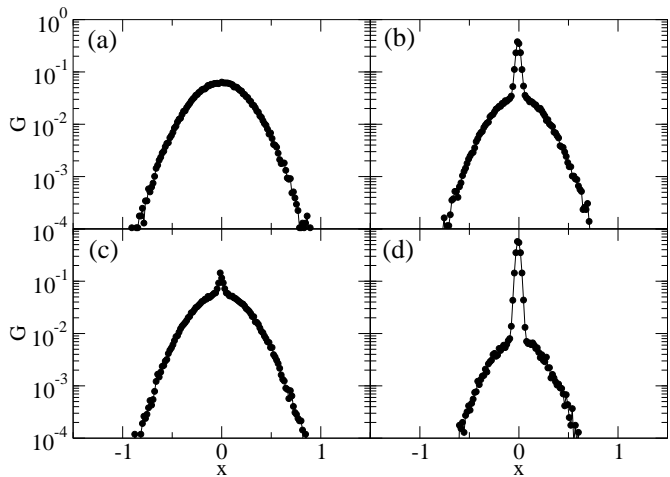}
\caption{ 
The self part of the van Hove correlation distribution function
for the $x$ direction for $f_{p} = 2.0$, $T=1.0$, and 
(a) $n_{p} = 0.0$, (b) $n_{p} = 0.5$, (c) $n_{p} = 0.1$, 
and (d) $n_{p} = 0.9$. 
}
\end{figure}

In Fig.~13(a) we plot the self part of the 
van Hove correlation function $G$ computed along the $x$ direction
for a system with $n_{p} = 0.0$ and $T = 1.0$.  This system is in 
the liquid state, since the initial disordering temperature 
for $n_p=0$ occurs at $T = 0.6$. 
Here, we find a single Gaussian distribution, indicating isotropic
transport. In Fig.~13(b) we show $G$ for $T  = 1.0$ and $n_{p} = 0.5$, 
where the system is in the heterogeneous regime. 
Two features appear. There is
a peak near $x = 0.0$ which corresponds to a group of
particles that are much less mobile
during the time frame of the measurement. 
These are the colloids located at the
pinning sites, which undergo only a small amount of motion within the pinning
sites, producing a peak in $G$ with very narrow width.
Superimposed on this narrow peak is
a second, wider Gaussian distribution that corresponds to the 
colloids in the liquidlike regions which can travel a 
much larger distance.
We plot $G$ for $T=1.0$ and $n_p=0.1$ in Fig.~13(c).
There is a much smaller narrow peak at $x =0$, consistent with the
fact that a much smaller fraction of the colloids is pinned and thus
fewer colloids have low mobility.
In Fig.~13(d) we show the case for $n_{p} = 0.9$ at the
same temperature, where the peak at $x = 0.0$ 
is high and the wide part of the distribution is smaller.  
For all $n_{p}$ 
at $T > 3.0$, the distribution function appears
very similar to the one shown in Fig.~13(a), with increased width.  

\section{Conclusion}

We have studied the dynamic and topological heterogeneities in a 
2D system of interacting particles with pinning.
The number of colloids is fixed and is commensurate with a triangular
pinning substrate. By shutting off a fraction of the pinning sites
randomly, 
we can control the amount of heterogeneous motion.
For sufficiently strong pinning, there can be a two step melting
process in which the colloids in the unpinned regions melt first
followed by the colloids in the pinned regions.
The two step melting appears as a double peak in the derivative of
the density of six-fold coordinated particles.  
The motion at temperatures between the two melting transitions is
heterogeneous, and the topological defects are associated with
the more mobile unpinned regions. The creation
and annihilation of the topological defects
in the mixed liquid-solid phase shows a prominent $1/f^{\alpha}$
power spectrum.
The noise power reaches a much higher value
in the mixed phase than in a system with no pinning.
For weaker disorder, the two step melting is lost. 
Signatures of the heterogeneous motion can also be observed in the
van Hove correlation function, 
which is composed of two overlapping Gaussian distributions.
In the high temperature homogeneous phase, only a single Gaussian distribution
appears.
Our results suggest that the dynamically heterogeneous 
regions in which the colloids are more mobile may be associated 
with regions that contain a larger number of topological 
defects and which are locally molten.  We also predict that
the temporal fluctuations of the density of defects in regions that
show dynamical heterogeneities will have a $1/f^{\alpha}$ noise signature, and
that at higher temperatures where the heterogeneities are lost, the system will
have a white noise spectrum.  
This is the first study of dynamical heterogeneities in 
a system with quenched disorder. 

We thank M. Dykman for helpful discussions.
This work was supported by the U.S. DoE under Contract No.
W-7405-ENG-36.

\end{document}